\documentclass[preprint]{aastex}
\shorttitle{Relation between EIT Waves and CMEs}
\slugcomment{Submitted to ApJL}
\shortauthors{Chen}
\begin{document}

\title{The Relation between EIT Waves and Coronal Mass Ejections}
\author{P. F. Chen}
\affil{Department of Astronomy, Nanjing University, Nanjing 210093,
China; chenpf@nju.edu.cn}

\begin{abstract}
More and more evidence indicates that ``EIT waves" are strongly related
to coronal mass ejections (CMEs). However, it is still not clear how the
two phenomena are related to each other. We investigate a CME event on
1997 September 9, which was well observed by both EUV imaging telescope
(EIT) and the high-cadence MK3 coronagraph at Mauna Loa Solar
Observatory, and compare the spatial relation between the ``EIT wave"
fronts and the CME leading loops. It is found that ``EIT wave" fronts
are co-spatial with the CME leading loops, and the expanding EUV
dimmings are co-spatial with the CME cavity. It is also found that the
CME stopped near the boundary of a coronal hole, a feature common to
observations of ``EIT waves". It is suggested that ``EIT waves"/dimmings
are the EUV counterparts of the CME leading loop/cavity, based on which
we propose that, as in the case of ``EIT waves", CME leading loops are
apparently-moving density enhancements that are generated by successive
stretching (or opening-up) of magnetic loops.
\end{abstract}


\section{Introduction}
As the largest-scale eruptive phenomenon on the Sun, coronal mass
ejections (CMEs) are often accompanied by many other phenomena, which
are visible in various wavelengths, such as solar flares, filament
eruptions, and ``EIT waves". Among them, ``EIT waves" are extremely
enigmatic. They have attracted wide attention and provoked a lot of
debate concerning their physical nature. The first reports of the
phenomenon appeared in \citet{dere97} and \citet{mose97}. \citet{thom98}
analyzed the famous 1997 May 12 event in detail using the data from the
EUV Imaging Telescope (EIT) on board the {\it Solar and Heliospheric
Observatory} ({\it SOHO}) spacecraft. In association with the CME event,
the running difference images of the EIT data showed that an almost
circular bright front, with an averaged intensity enhancement of $\sim
20\%$ \citep*{thom99}, propagates away from the source active region,
giving an impression of a wave. They were conventionally called ``EIT
waves" since they were discovered with the EIT telescope, though
sometimes they are referred to as ``coronal waves" \citep*{warm04,
attr07,trip07}.

As ``EIT waves" propagate outward, they are immediately followed by
expanding dimmings. Therefore, the two phenomena were proposed to result
from the same physical process, which was very controversial during the
past years \citep[e.g.,][]{chen08}. One debate concerns the driving
source of the wave propagation. It was quite often claimed that ``EIT
waves" are generated by the pressure pulse which may be a solar flare
\citep*[e.g.,][]{wu01,vrsn02}, whereas \citet{plun98}, \citet{dela99},
\citet{chen02}, and \citet{chen05} proposed that they are associated
with CMEs. \citet{bies02} studied 173 events and found that after
correcting for observational biases all EIT waves are associated
with CMEs. Noticing that about half of the ``EIT waves" observed from
1997 March to 1998 June were associated with small solar flares below C
class, \citet{cliv05} pointed out that it is hard to imagine that such
weak flares can produce global-scale perturbations, so that ``EIT waves"
should be preferentially associated with CMEs. As a complementary proof,
\citet{chen06} selected strong (M- and X-class) flares near solar
minimum that were not associated with CMEs, and found that none of them
produced ``EIT waves". Furthermore, \citet{chen06} showed that from the
same active region within the same day, a weaker flare with a CME was
associated with an ``EIT wave", however, another stronger flare without
a CME was not accompanied by an ``EIT wave". In addition, \citet{podl05}
and \citet{attr07} found that EIT waves rotate in senses determined by
the helicity of the CME source region, a feature not expected from
flare-induced waves. These pose strong evidence to support that ``EIT
waves" are purely related to CMEs.  However, the physical connection
between ``EIT waves" and CMEs is still not well established.
\citet{vero08} proposed that ``EIT waves" are fast-mode waves driven by
CME flanks, while \citet{chenf05} theoretically postulated that ``EIT
wave" fronts are the EUV signature of the CME leading loops, and
accordingly, the expanding EUV dimmings are the EUV signature of the CME
cavity. With the data analysis of the CME event on 1997 September 9,
this Letter aims to clarify the connection between ``EIT waves" and CMEs.

\section{Observations}

``EIT waves" are observed by EUV imaging telescopes like EIT, with a
field of view (FOV) of $\sim 1.5R_\sun$. Before the launch of the recent
STEREO satellites, CMEs were mostly observed by three coronagraphs (C1,
C2, and C3) in the Large Angle and Spectrometric Coronagraph (LASCO),
which are on board the {\it SOHO} spacecraft. The white-light
coronagraphs C2 and C3, which observe the coronal mass directly, have a
FOV beyond $2R_\sun$. Therefore, the comparison of the observations
between them and EIT relies on spatial extrapolation, which introduces
uncertainties.  The C1 coronagraph has a FOV from $1.1-3R_\sun$, which
overlaps with that of EIT. However, during its short lifetime, it was
often observing the corona with the forbidden lines, whose intensity
depends not only on the coronal density, but also on the temperature.
Therefore, their structure may not necessarily be co-spatial with the
white-light CME. In order to precisely determine the spatial relation
between ``EIT wave" fronts and CMEs, we study ground-based coronagraph
 white-light data with a FOV overlapping that of EIT.

\citet{thom09} compiled a catalog of 176 ``EIT wave" events that
occurred between 1997 March and 1998 June. In order to compare the ``EIT
wave" fronts and CMEs, we selected those ``EIT wave" events that have
fronts appearing above the solar limb. It is found that 38 cases in the
catalog are off-limb events. We then search for the white-light
data from the Mark-III K-Coronameter (MK3), which was installed at the
Mauna Loa Solar Observatory (MLSO). It turned out that only the 1997
September 9 event was well observed by the MK3 coronagraph. The ``EIT
wave" propagation was observed by the EIT instrument, and the CME was
observed by both the ground-based coronagraph MK3 and the space-borne
LASCO instrument.

The EIT instrument is a normal-incidence, multilayer EUV telescope
\citep{dela95}. It observes the full-disk solar corona, extending up to
$1.5R_\sun$ with a pixel size of $2.59^{\prime\prime}$. There are four
narrow bandpass EUV channels centered at 171, 195, 284, and 304 {\AA},
which selectively observe spectral lines formed by \ion{Fe}{9/X},
\ion{Fe}{12}, \ion{Fe}{12}, and \ion{He}{2}, respectively. The
\ion{Fe}{12} 195 {\AA} images used in
this paper have a cadence of $\sim16$ min.  The \ion{Fe}{12} emission
line exhibits a peak emission near 1.5 MK.  The LASCO instrument
consists of a set of three coronagraphs, i.e., C1, C2, and C3, with
overlapping and concentric FOV. C2 and C3 are traditional white-light
coronagraphs that observe Thomson-scattered visible light through a
broadband filter \citep{brue95}. The MK3 coronagraph at MLSO began
observations in 1980 \citep{macq83}. It measures polarization
brightness of photospheric radiation scattered by free electrons in the
lower corona, with a FOV of $1.12-2.45R_\sun$, a pixel size of
$20^{\prime\prime}$, and a cadence of 3 min.

\section{Results}\label{sec3}

\begin{figure}
\epsscale{1.10}
\plotone{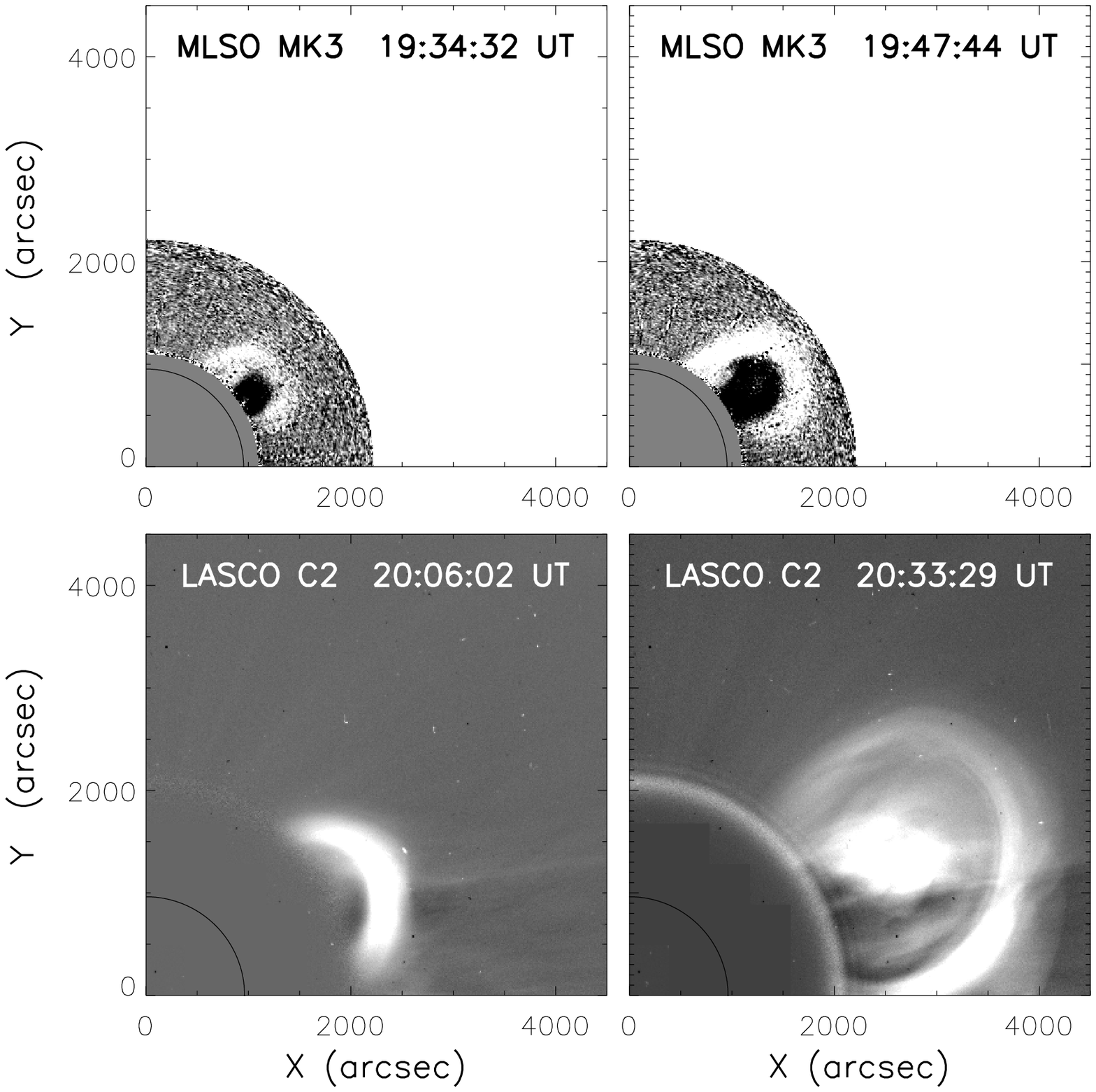}
\caption{Evolution of the base-difference white-light intensity of the
CME on 1997 September 9, where the black arcs mark the solar limb. {\it
Upper panels}: images observed by MK3 coronagraph in MLSO. {\it Lower
panels}: images observed by LASCO C2 coronagraph.
\label{f1}}
\end{figure}

The leading loop of the CME was visible in the FOV of the MK3
coronagraph
from 19:34:32 UT on 1997 September 9. The data before that were not
of high enough quality to clearly show the structure of the CME. The
upper panels of Figure \ref{f1} display the white-light base-difference
images of the CME, which were observed by MK3. The pre-event intensity
map at 19:04:11 UT is chosen as the base image that is subtracted from
the later images. The high-cadence observations of MK3 indicated that
the CME leading loop accelerated from 146$\pm$36 km s$^{-1}$ at
19:34:32 UT to 366$\pm$36 km s$^{-1}$ at 19:47:44 UT, after which the
top of the leading loop went out of the FOV of the MK3. At 20:06:02 UT, the
CME began to be visible in the FOV of the LASCO C2 coronagraph at
$2.6R_\sun$, as seen from the lower panels of Figure \ref{f1}. At
20:33:29 UT, the CME leading loop moved to a heliocentric distance of
$5.5R_\sun$ in the plane-of-the-sky. The averaged radial speed of the
CME propagation is estimated to be 726$\pm$20 km s$^{-1}$ in the FOV of
LASCO C2, which is almost twice that measured in the FOV of the MK3.

\begin{figure}
\epsscale{1.0}
\plotone{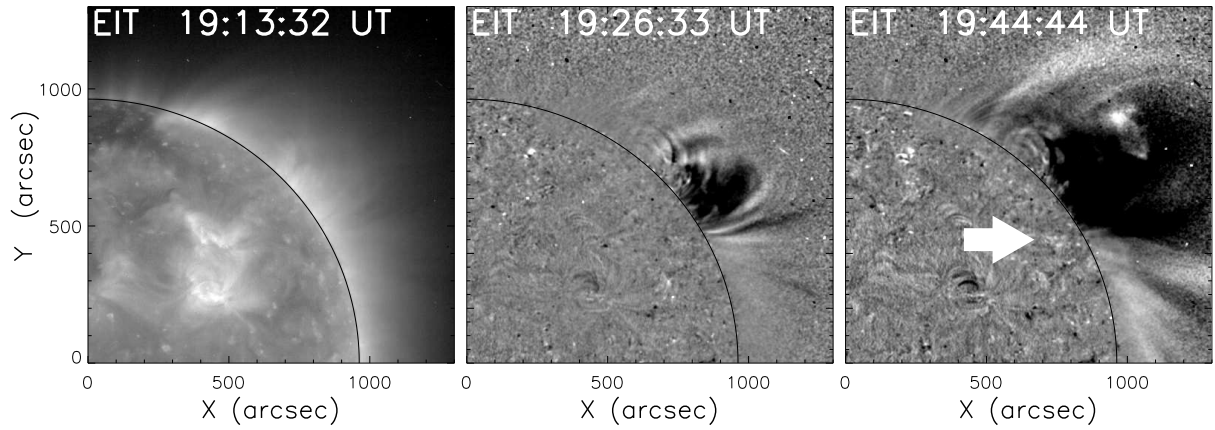}
\caption{EIT 195{\AA} images showing the evolution of the ``EIT wave"
event. {\it Left}: Pre-event image at 19:13:32 UT. {\it Middle}:
Base difference image at 19:26:33 UT. {\it Right}: Base difference image
at 19:44:44 UT. White arrow shows weak brightening of the ``EIT wave"
visible on the solar disk.
\label{f2}}
\end{figure}

As the CME lifted off, no flare-like brightening was visible on the
solar disk, inferring that the source region of this CME event was
located behind the solar limb. However, the full-disk imager, EIT,
detected the propagation of an ``EIT wave", as revealed by the middle
and right panels of Figure \ref{f2}. In this figure, the left panel
shows the pre-event \ion{Fe}{12} 195 {\AA} image, and the middle and
right panels depict the evolution of the base-difference 195 {\AA}
images. It is seen that the main part of the ``EIT wave" was above the
limb, extending beyond the FOV of EIT, with some weak brightenings on
the solar disk as indicated by the white arrow in the right panel. As
revealed by the left panel of Figure \ref{f2}, a polar coronal hole
existed on the northern side of the ``EIT wave". The northern leg of
the ``EIT wave" was approaching the coronal hole from 19:26:33 UT to
19:44:44 UT, after which it stopped near the boundary of the coronal
hole.

Owing to the high cadence of the MK3 observations, the spatial positions
of the white-light CME and the ``EIT wave" can be compared directly when
the two phenomena were observed almost simultaneously. The white-light
images of the CME at 19:44:27 UT and 20:00:49 UT are displayed in the
upper panels of Figure \ref{f3}, and the corresponding EIT 195 {\AA}
images at 19:44:44 UT and 20:00:14 UT are shown in the lower panels. For
clearness, the outlines of the white-light CME leading loops (defined by
eye)  in the upper left and right panels are superimposed on the EIT
\ion{Fe}{12} intensity map in the lower left and right panels,
respectively, as represented by the thick black lines. It is seen that
the ``EIT wave" fronts above the limb are almost co-spatial with the
leading loop of the CME, with slight differences in the detailed
structures. Accordingly, the EIT dimmings are co-spatial with the
white-light cavity of the CME. Note that there is a small time
difference in the white-light and EUV observations ($\sim$35 s).
Considering that the top of the CME leading loop was moving with a speed
of 366$\pm$36 km s$^{-1}$ in the FOV of the MK3, such a time difference
corresponds to a spatial shift of $18^{\prime\prime}$, which is below
the spatial resolution of the MK3 coronagraph.

\section{Discussions}

\subsection{Spatial Relation between ``EIT Waves" and CMEs}

More and more evidence tends to support that ``EIT waves" are related to
CMEs, rather than solar flares. However, exactly how ``EIT waves" are
related to CMEs is still not clear. The simultaneous observations of an
``EIT wave" in \ion{Fe}{12} 195 {\AA} and a white-light CME in the 1997
September 9 event provide a precious opportunity to tackle this question.

\begin{figure}
\epsscale{1.0}
\plotone{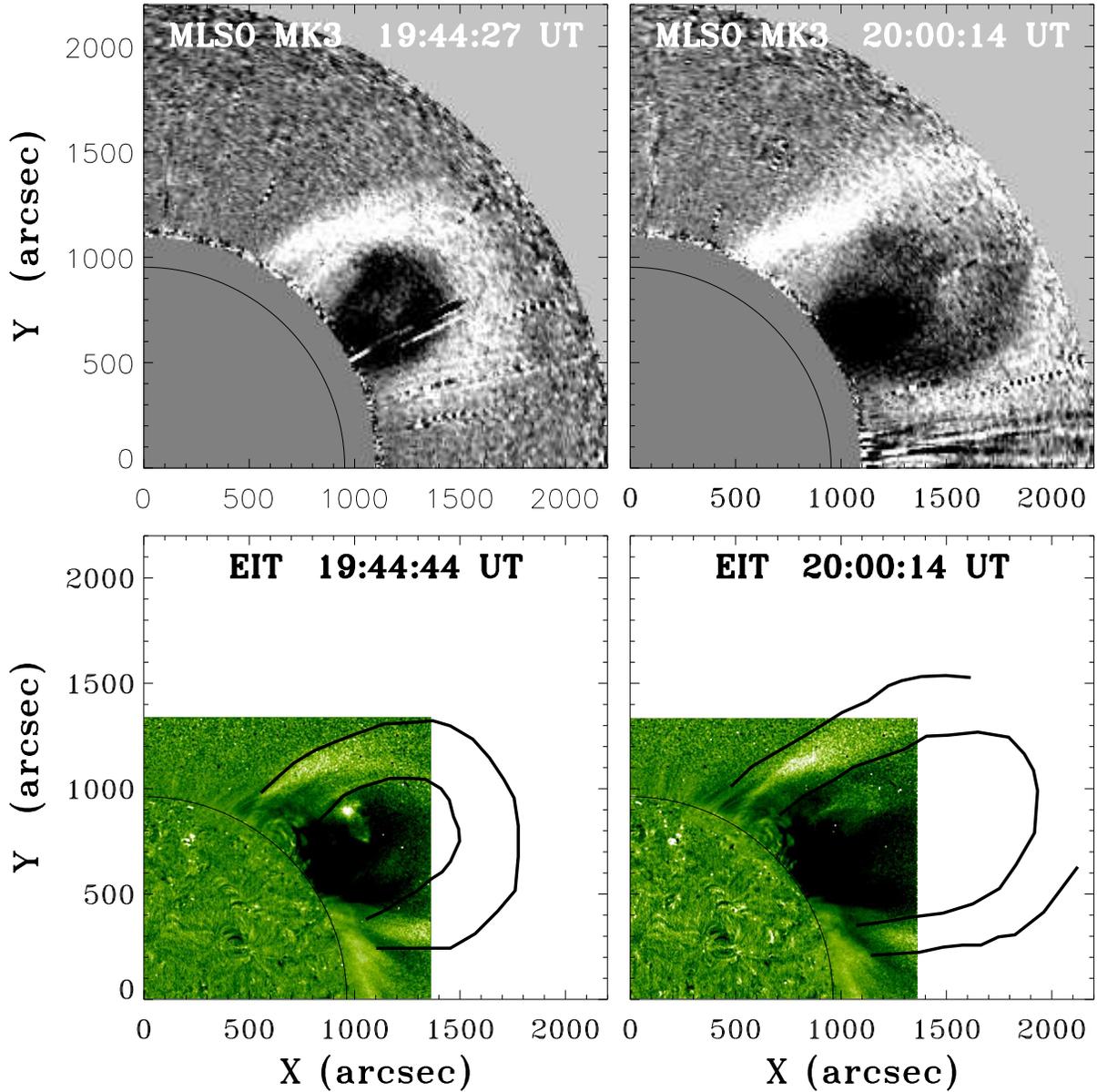}
\caption{Spatial comparison of the CME leading loop and the ``EIT wave"
front. {\it Upper panels}: Two snapshots of the white-light CME
eruption observed by the MLSO MK3 coronagraph at 19:44:27 UT and
20:00:49 UT. {\it Lower panels}: The \ion{Fe}{12} 195 {\AA} images
of the ``EIT wave" propagation observed by the EIT instrument at
19:44:44 UT and 20:00:14 UT.
\label{f3}}
\end{figure}

The source region of the eruption was located behind the solar limb,
hence the ``EIT wave" was observed as a limb event, which allows the
direct comparison with coronagraph data. Our analysis in Section
\ref{sec3} (see Figure \ref{f3}) indicates that the ``EIT wave" fronts
were almost co-spatial with the leading loops of the CME during the
eruption, and the expanding EUV dimmings, which immediately followed the
``EIT wave" fronts, were co-spatial with the dark cavity of the CME.

The brightening of the ``EIT wave" fronts in \ion{Fe}{12} 195 {\AA}
might be due to temperature variation and/or density enhancement
\citep{thom99}. Therefore, it is controversial whether the brightening
is mainly contributed by temperature variation \citep[e.g., due to Joule
heating,][]{dela08} or by density enhancement \citep{will99}.
Considering that ``EIT waves" are observed simultaneously in several EUV
lines that have different formation temperatures \citep{will99,long08},
it is believed that the ``EIT wave" brightening is mainly due to density
enhancement, though the adiabatic compression may increase the plasma
temperature to some extent, resulting in some differences between
detailed features in different lines \citep{will99,chenf05}. The
approximate co-spatiality of ``EIT wave" fronts and white-light CME
leading loops, as revealed by Figure \ref{f3}, provides direct evidence
that ``EIT wave" brightenings are mainly contributed by density
enhancement, since the white-light enhancement of the CME leading loops
is produced by the increased coronal density only. Therefore, we
conclude that ``EIT wave" fronts are mainly due to density enhancement,
and they are the EUV signatures of the CME leading loops, as we
theoretically proposed in \citet{chenf05}. It is also inferred that,
similar to the frequently assumed dome-like shape of CME leading loops,
the ``EIT wave" front should also be dome-like, and the circular ``EIT
wave" fronts, sometimes observed on the solar disk, are just a
projection of the three-dimensional dome-like structure, whose skirt is
much brighter than the top of the dome.

\subsection{Nature of CME Leading Loop}

The physics behind the formation of CME leading loops is still not
clear. It is often taken for granted that we observe coronal plasma
embedded in the erupting magnetic loops. However, with UV spectral
observations of halo CMEs, \citet{ciar06} found that the CME fronts
show Doppler shifts significantly smaller than their apparent
velocity obtained with white-light coronagraphs, suggesting that CME
leading loops (at least for the halo events), might be fast-mode shocks
rather than being plasma carried outward by erupting magnetic loops.
The co-spatiality of ``EIT waves" and CME leading loops found in this
Letter could also shed light on the nature of the CME leading loops.

``EIT waves" are often considered to be fast-mode magnetoacoustic waves
in the corona \citep[e.g.,][]{wang00,wu01,vrsn02,warm04,grec08,pomo08}.
However, the wave model cannot explain the following features of ``EIT
waves" \citep[e.g., see][for details]{will07,chen08}: (1) the ``EIT wave"
velocity is significantly smaller than those of Moreton waves. The
latter are well established to be due to fast-mode waves in the corona;
(2) The ``EIT wave" velocities have no correlation with those of type II
radio bursts \citep{klas00}; (3) The ``EIT wave" fronts may stop when
they meet with magnetic separatrices \citep{dela99}; (4) The ``EIT wave"
velocity may be below 100 km s$^{-1}$ \citep[e.g., the Fig. 3
of][]{long08}, which is even smaller than the sound speed in the corona.
These strange features provoked \citet{dela99} to relate ``EIT waves" to
magnetic restructuring during CMEs. With MHD numerical simulations, Chen
et al. (\citeyear{chen02, chen05}) identified the ``EIT wave" features
to correspond to stretching of the magnetic field. The ``EIT wave"
disturbance was observed well behind the fast-mode piston-driven shock
waves during CME eruptions. Chen et al. (\citeyear{chen02, chen05})
proposed that ``EIT waves" are apparently-moving density enhancements,
which are actually produced by successive stretching (or opening-up) of
closed field lines, rather than being real waves. The model can account
for the main characteristics of ``EIT waves", such as the low velocity,
their diffuse fronts, the stationarity near magnetic separatrices, and
found support in observations \citep[e.g.,][]{harr03}. The co-spatiality
of ``EIT waves" and CME leading loops found in this paper would infer
that CME leading loops are also generated by the successive stretching
of overlying magnetic loops. As illustrated by Figure 4, as the core
structure, e.g., a magnetic flux rope, erupts, the resulting
perturbation propagates outward in every direction, with a probability
of forming a piston-driven shock as indicated by the pink lines.
However, different from a pressure pulse, the erupting flux rope
continues to push the overlying magnetic field lines to move outward, so
that the field lines are stretched outward one by one. For each field
line, the stretching starts from the top, e.g., point A for the first
magnetic line, and then is transferred down to the leg (point D) with
the Alfv\'en speed. The deformation at point A is also transferred
upward to point B of the second magnetic line with the fast-mode speed.
Such a deformation would also be transferred down to its leg (point E)
with the local Alfv\'en speed, by which the entire second magnetic line
is stretched up. The stretching at any part of the magnetic field lines
compresses the coronal plasma on the outer side, producing density
enhancements. All the newly formed density enhancements form a pattern
({\it green}), which is observed as the CME leading loop. Similar to
``EIT wave" fronts, the legs of the CME leading loop separate initially,
and may stop when they meet with magnetic separatrices such as the
boundary of coronal holes. This is why CMEs generally maintain a fixed
angular span in their later stages. At the same time, as the field lines
are stretched outward, the enveloped volume increases, resulting in
coronal dimmings (or the dark cavity) behind the CME leading loop.

To conclude, with the simultaneous observations of an ``EIT wave" in
\ion{Fe}{12} 195 {\AA} and a CME in white light, we found that ``EIT
wave" fronts are co-spatial with CME leading loops, and accordingly, 
expanding EUV dimmings are co-spatial with the CME cavity. We postulate
that the CME leading loops may have the same formation mechanism and
physical nature as ``EIT waves", i.e., they are the apparently-moving
density enhancements that are generated by successive stretching (or
opening-up) of magnetic loops. It is noted that in other ``EIT
wave" models like the reconnection model of \citet{attr07}, the ``EIT
wave" front is also expected to be co-spatial with the CME leading loop.
The difference between these models should be explored further.

\acknowledgments

The author thanks C. Fang and the referee for the constructive
suggestions, and F. Gu for the help of data analysis. The research is
supported by the Chinese foundations 2006CB806302 and NSFC (10403003,
10221001, and 10333040).
{\it SOHO} is a project of international cooperation between ESA and
NASA. We are grateful to the MLSO team for making their data available.

\clearpage

\clearpage

\clearpage

\clearpage

\begin{figure}
\epsscale{1.}
\plotone{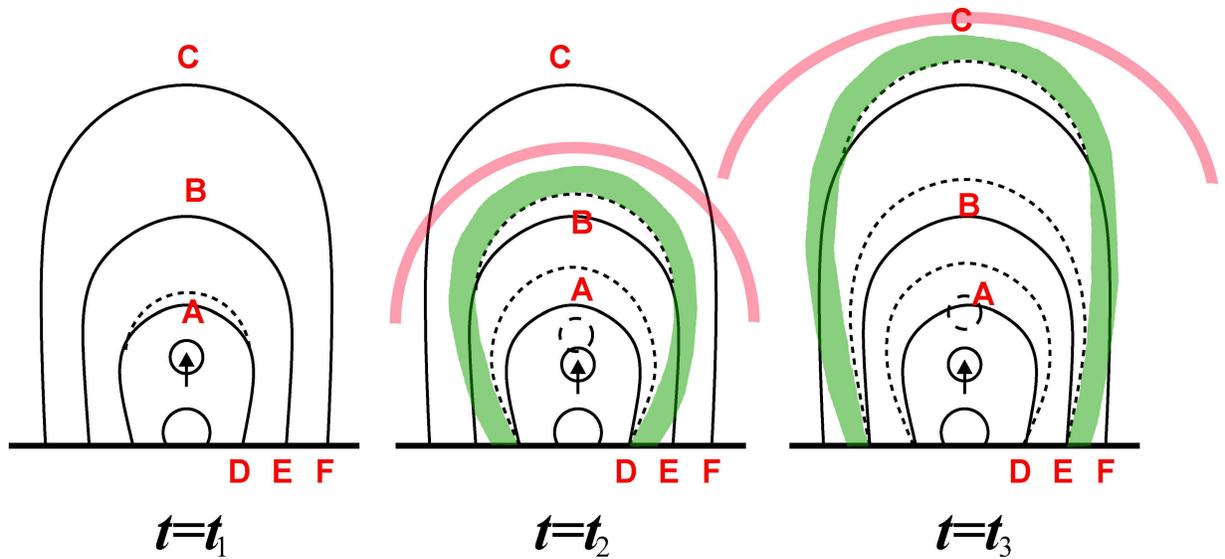}
\caption{A schematic sketch of the formation mechanism of CME leading
loops, where the CME leading loop ({\it green}) are apparently-moving
density enhanced structure that is generated by the successive
stretching of magnetic field lines as the erupting core structure, e.g.,
a flux rope, continues to push the overlying field lines to expand
outward. The piston-driven shock is shown as pink lines.
\label{f4}}
\end{figure}

\end{document}